\documentclass[12pt]{article}
\usepackage{epsfig,amsmath,amsfonts,amssymb,makeidx,ifthen}
\makeatletter
\@addtoreset{equation}{section}
\makeatother

\newcommand{\ret}{\nonumber\\}
\newcommand{\bigno}{\par\bigskip\noindent}
\newcommand{\abs}[1]{\left|#1\right|}

\newcommand{\sbkt}[1]{\langle#1\rangle}

\newcommand{\sumtwo}[2]%
{\mathop{\sum_{#1}}_{#2}}
\newcommand{\sumthree}[3]%
{\mathop{\mathop{\sum_{#1}}_{#2}}_{#3}}
\newcommand{\sumfour}[4]%
{\mathop{\mathop{\mathop{\sum_{#1}}_{#2}}_{#3}}_{#4}} 
\newcommand{\suptwo}[2]%
{\mathop{\sup_{#1}}_{#2}}
\newcommand{\supthree}[3]%
{\mathop{\mathop{\sup_{#1}}_{#2}}_{#3}}
\newcommand{\supfour}[4]%
{\mathop{\mathop{\mathop{\sup_{#1}}_{#2}}_{#3}}_{#4}} 
\newcommand{\inftwo}[2]%
{\mathop{\inf_{#1}}_{#2}}
\newcommand{\infthree}[3]%
{\mathop{\mathop{\inf_{#1}}_{#2}}_{#3}}
\newcommand{\inffour}[4]%
{\mathop{\mathop{\mathop{\inf_{#1}}_{#2}}_{#3}}_{#4}} 

\newcommand{\calL}{{\cal L}}

\newcommand{\calS}{{\cal S}}
\newcommand{\calT}{{\cal T}}

\newcommand{\rhoss}{\rho_\mathrm{st}}
\newcommand{\Gs}{{\Gamma}_\mathrm{s}}
\newcommand{\Hs}{H_\mathrm{s}}
\newcommand{\Hstat}{H_\mathrm{stat}}
\newcommand{\Te}{\calT}
\newcommand{\Tei}{\calT^{-1}}
\newcommand{\ofs}[1]{(#1)_\mathrm{s}}
\newcommand{\Ss}{\calS_\mathrm{s}}
\newcommand{\Di}{\mathit{\Delta}}
\newcommand{\bsp}{\boldsymbol{p}}
\newcommand{\bsr}{\boldsymbol{r}}
\newcommand{\bsbeta}{\boldsymbol{\beta}}
\newcommand{\betaeq}{\beta_\mathrm{eq}}

\newcommand{\bsmu}{\boldsymbol{\mu}}
\newcommand{\Sex}{\Theta_\mathrm{ex}}
\newcommand{\pss}{\varphi_\mathrm{st}}

\newcommand{\sss}{\mathrm{st}}
\newcommand{\eqs}{\mathrm{eq}}
\newcommand{\Gammas}{\Gamma^*}
\newcommand{\Upsilons}{\Upsilon^*}
\newcommand{\gammas}{\gamma^*}
\newcommand{\nus}{\nu^*}
\newcommand{\taus}{\tau^*}
\newcommand{\barsigma}{\overline{\sigma}}
\newcommand{\phizero}{\varphi^\mathrm{(0)}}

\newcommand{\phic}{\varphi^\mathrm{(c)}}

\newcommand{\Ns}{N_\mathrm{s}}
\newcommand{\dd}{d}
\newcommand{\dt}{{\dd}t}
\newcommand{\dG}{\dd{\Gamma}}
\newcommand{\dGs}{\dd\Gs}
\newcommand{\Hb}{H^\mathrm{(b)}}
\newcommand{\Hc}{H^\mathrm{(c)}}

\newcommand{\Ms}{M_\mathrm{s}}
\makeatletter
\long\def\@makecaption#1#2{{\small
\advance\leftskip1cm
\advance\rightskip1cm
\vskip\abovecaptionskip
\sbox\@tempboxa{#1: #2}%
\ifdim \wd\@tempboxa >\hsize
 #1: #2\par
\else
\global \@minipagefalse
\hb@xt@\hsize{\hfil\box\@tempboxa\hfil}%
\fi
\vskip\belowcaptionskip}}
\makeatother
\begin{document}
\noindent
{\bf\LARGE Representation of nonequilibrium steady states in  large  mechanical systems}
\par\bigskip

\noindent
Teruhisa S. Komatsu\footnote{
Department of Physics, Gakushuin University, Mejiro, Toshima-ku, Tokyo 171-8588, Japan
}, Naoko Nakagawa\footnote{
College of Science, 
 Ibaraki University, Mito, Ibaraki 310-8512, Japan
}, Shin-ichi Sasa\footnote{
Department of Pure and Applied Sciences, The University of Tokyo,
 Komaba, Meguro-ku, Tokyo 153-8902, Japan
}, and Hal Tasaki${}^1$

\bigskip

\begin{abstract}
Recently a novel concise representation of the probability distribution of heat conducting nonequilibrium steady states was derived.
The representation is valid to the second order in the ``degree of nonequilibrium'', and has a very suggestive form where the effective Hamiltonian is determined by the excess entropy production.
Here we extend the representation to a wide class of nonequilibrium steady states realized in classical mechanical systems where baths (reservoirs) are also defined in terms of deterministic mechanics.
The present extension covers such nonequilibrium steady states with a heat conduction, with particle flow (maintained either by external field or by particle reservoirs), and under an oscillating external field.
We also simplify the derivation and discuss the corresponding representation to the full order.
\end{abstract}

\tableofcontents

\section{Introduction}
To construct a statistical mechanics that applies to nonequilibrium states is one of the most challenging unsolved problems in theoretical physics.
See \cite{ST} and references therein.
By a statistical mechanics, we mean a universal theoretical
framework that enables one to precisely characterize 
 states of a given system, and to compute
(in principle) arbitrary macroscopic quantities.
The canonical distribution for equilibrium states, in which the probability of observing a microscopic state $\gamma$ is  given by $\rho(\gamma)\propto\exp[-\beta\,H(\gamma)]$, is a paradigm for a statistical mechanics.

It is, however, quite unlikely that there exists a statistical mechanics  that applies to {\em any}\/ nonequilibrium systems.
A much more modest (but still extremely ambitious)
goal is to look for a theory that applies to nonequilibrium steady
states, which have no macroscopically observable time dependence but have macroscopic flow of energy or material.
There may be a chance that 
probability distributions for nonequilibrium steady states can be 
obtained from a general principle that is
analogous to the  equilibrium statistical mechanics.

If we restrict ourselves to those models with extremely small
``order of nonequilibrium'', the linear response theory provides us
with a more or less satisfactory answer.
See, for example, \cite{KuboTodaHashitsume85}. 
One can represent steady state distribution and various physical quantities
by using time-dependent correlation functions in the corresponding 
equilibrium state.
But the restriction to the linear order is unsatisfactory at least from a purely theoretical point of view.
It is highly desirable and challenging to obtain similar expressions which works beyond the linear response regime.

In fact, formal expressions of the steady state distribution which are exact to full order were  derived and discussed, for example, by McLennan \cite{McLennan90}, Zubarev \cite{Zubarev74}, and  Kawasaki and Gunton \cite{KawasakiGunton73}.
See \eqref{e:exact} for an example of such expressions.
But such expressions, as they are, are too formal and do not directly provide us with  meaningful information about the nature of  nonequilibrium steady states.
See section~\ref{s:comp}.
Such expressions were indeed used as starting points of further explorations of nonequilibrium steady states.

In a recent progress (which is of course closely related to previous works that we have mentioned) in nonequilibrium physics, deep implications of the microscopic time-reversal symmetry were revealed \cite{ECM,GC,Kurchan,Maes,Crooks,J,MN}.
It was shown that seemingly simple symmetry has rich and meaningful consequences including the fluctuation theorem and the nonequilibrium work theorem.

In this connection, two of us (Komatsu and Nakagawa) studied the heat conducting nonequilibrium steady state realized in a system attached to multiple heat baths, and obtained a novel concise representation of the steady state distribution \cite{KNPRL}.
The representation is written in terms of the (excess) entropy production at the heat baths, and is correct to the second order in the heat current.

Although the result of \cite{KNPRL} is also based on the time-reversal symmetry, it stands out from the previous works in the following two points.
First this result directly addresses the question about the precise form of the probability distribution in nonequilibrium steady states.
Secondly \cite{KNPRL}  presents a non-exact result which is valid up to the second order in the heat current.
We regard this restriction as a merit rather than a fault.
By looking only at exact relations, one is tempted to be satisfied with rather formal results which do not focus strongly on desired physics of nonequilibrium states.
Well controlled result for small ``degree of nonequilibrium'' (like the one in \cite{KNPRL}) may suggest various
nontrivial natures of nonequilibrium steady states.

We expect that this suggestive representation can be a starting point of further developments of nonequilibrium physics.
In fact we have made use of this representation to derive thermodynamic relations for nonequilibrium steady states \cite{KNSTshort}.

\bigskip

In the present paper, we do not go into applications of the representation of \cite{KNPRL}, but rather focus on its more basic aspects.
We shall discuss some extensions, and also present an efficient derivation of the representation.

In \cite{KNPRL}, the representation was derived for stochastic processes.
Since the essence of the representation is the microscopic time-reversal symmetry, this restriction is by no means essential.
Here we shall present a derivation of the representation for general models of nonequilibrium steady states which are described entirely in terms of deterministic classical mechanics.
More precisely, we design the whole system (including the heat or particle reservoirs) using deterministic mechanics.
By letting the whole system evolve for a sufficiently long time, we get a nonequilibrium steady state (in a small part of the whole system).
We can treat nonequilibrium steady states with a heat conduction or with a particle current (maintained either by non-conservative external force or by particle reservoirs with different chemical potentials).
We can also treat a nonequilibrium state which is maintained by an oscillating external force.

In \cite{KNPRL}, only the lowest order of the representation was discussed in detail (because of the limitation of the space).
We here discuss the corresponding formal representation which is valid to the full order in the ``order of nonequilibrium.''
The derivation here is essentially the same as that in \cite{KNPRL}, but we have refined the argument so that to make it as transparent as possible.

\bigskip

The present paper is organized as follows.
In section~\ref{s:Setting}, we carefully describe our setting, and how one can realize a nonequilibrium steady state in a system described by deterministic (Hamiltonian or Newtonian) mechanics.
In section~\ref{s:Representaion}, we introduce the notion of excess entropy production, and describe the main representation.
In section~\ref{s:Derivation}, we derive the representation.
Finally, in section~\ref{s:paricleflow}, we formulate nonequilibrium steady states with particle flow, and extend the representation.

\section{Setting}
\label{s:Setting}
In the present section, we carefully describe the problem that we study.
In short, we  construct deterministic classical mechanical systems of many particles which faithfully model typical situations where we expect to have nonequilibrium steady states.

\subsection{States and static Hamiltonians}
We consider a situation where a ``system'' is attached to $n$ large heat baths with different temperatures.
We shall model the whole system as a classical mechanical system which consists of $n+1$ distinct parts.
The first part is the system\footnote{
In what follows ``system'' always means the first part.
The collection of the $n+1$ parts is referred to as the ``whole system''.
} while the latter are the heat baths.
We assume that the different parts do not exchange particles, but 
the system and each heat bath may exchange energy.

By $\Gs=(\bsr_1^{(\mathrm{s})},\ldots,\bsr_{N_\mathrm{s}}^{(\mathrm{s})};\bsp_1^{(\mathrm{s})},\ldots,\bsp_{N_\mathrm{s}}^{(\mathrm{s})})$
 we collectively denote the coordinates and momenta of the system, which contains  $N_\mathrm{s}$ particles,
 and by $\Gamma_i=(\bsr_1^{(i)},\ldots,\bsr_{N_i}^{(i)};\bsp_1^{(i)},\ldots,\bsp_{N_i}^{(i)})$
 the coordinates and momenta of the $i$-th heat bath, which contains $N_i$ particles.
By $\Ss$ and $\calS_i$, we denote the phase spaces of the system and the $i$-th bath, respectively.
$\Gamma=(\Gs,\Gamma_1,\ldots,\Gamma_n)$ denotes the coordinates and momenta of the whole system.
The corresponding total phase space is  $\calS=\Ss\times\calS_1\times\cdots\times\calS_n$.
Finally $\dGs, \dG_i$ and  $\dG$ denote the  Lebesgue measures on $\Ss, \calS_i$ and $\calS$, respectively.
For any state $\Gamma=(\Gs,\Gamma_1,\ldots,\Gamma_n)\in\calS$, we denote by
 $\Gamma^*=(\Gamma^*_\mathrm{s},\Gamma^*_1,\ldots,\Gamma^*_n)\in\calS$ its time reversal, namely,
 the state obtained by reversing all the momenta, e.g., $(\bsr;\bsp)^*=(\bsr;-\bsp)$.

\begin{figure}
\centerline{\epsfig{file=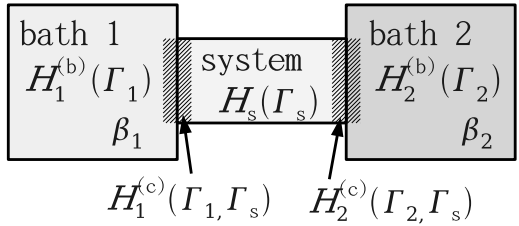,width=7cm}}
\caption[dummy]{
The whole system consists of a ``system'' on which we shall focus  and several heat baths.
Hamiltonians $\Hs$, $\Hb_i$, and  $\Hc_i$ describe the ``system'', the $i$-th bath, and the coupling between the ``system'' and the $i$-th bath, respectively.
We also write $H_i(\Gamma)=\Hb_i(\Gamma_i)+\Hc_i(\Gamma_i,\Gs)$.
}
\label{f:system}
\end{figure}

The static Hamiltonian of the whole system is written as
\begin{equation}
\Hstat(\Gamma)=\Hs(\Gs)+\sum_{i=1}^n\{\Hb_i(\Gamma_i)+\Hc_i(\Gamma_i,\Gs)\},
\label{e:Hstat}
\end{equation}
where $\Hs(\Gs)$ is the Hamiltonian for the system,
$\Hb_i(\Gamma_i)$ is that for the $i$-th bath, and $\Hc_i(\Gamma_i,\Gs)$ describes the coupling between the $i$-th bath and the system.
See Fig.~\ref{f:system}.
In many cases, we use the total Hamiltonian for the $i$-th bath
\begin{equation}
H_i(\Gamma)=\Hb_i(\Gamma_i)+\Hc_i(\Gamma_i,\Gs).
\label{e:Hidef}
\end{equation}

We assume that all the Hamiltonians have time-reversal symmetry, i.e., $\Hstat(\Gamma)=\Hstat(\Gamma^*)$,
 $\Hs(\Gs)=\Hs(\Gamma^*_\mathrm{s})$, and $H_i(\Gamma)=H_i(\Gamma^*)$ for $i=1,\ldots,n$.
 A typical choice is 
\begin{gather}
\Hs(\Gs)=\sum_{j=1}^{N_\mathrm{s}}
\frac{|\bsp_{j}^{(\mathrm{s})}|^2}{2m_j^{(\mathrm{s})}}
+V_\mathrm{s}(\bsr_1^{(\mathrm{s})},\ldots,\bsr_{N_\mathrm{s}}^{(\mathrm{s})}),
\label{e:HsE}
\\
\Hb_i(\Gamma_i)=
\sum_{j=1}^{N_i}
\frac{|\bsp_{j}^{(i)}|^2}{2m_j^{(i)}}
+V_i(\bsr_1^{(i)},\ldots,\bsr_{N_i}^{(i)}),
\label{e:HbE}
\\
\Hc_i(\Gamma_i,\Gs)=V_{i,\mathrm{s}}(\bsr_1^{(i)},\ldots,\bsr_{N_i}^{(i)};\bsr_1^{(\mathrm{s})},\ldots,\bsr_{N_\mathrm{s}}^{(\mathrm{s})}),
\label{e:HcE}
\end{gather}
where the potential $V_\mathrm{s}$ and $V_i$ represent both the external single-body forces (such as those from the walls) and the interaction between the particles.
The potential $V_{i,\mathrm{s}}$ represents the interaction between the particles in the $i$-th bath and in the system.

\subsection{Time evolution}
We denote by $\Te:\calS\to\calS$ the time evolution map of the whole system from time $t=0$ to $t=T$.
More precisely if the state at $t=0$ is $\Gamma$, then the state at time $t=T$ is $\Te(\Gamma)$.
By $\ofs{\Te(\Gamma)}\in\Ss$, we denote the state (i.e., coordinates and momenta) of the system in  $\Te(\Gamma)$.
(More generally $\ofs{\Gamma}\in\Ss$ denotes the projection of $\Gamma\in\calS$ onto $\Ss$.)
We do not make the $T$-dependence explicit since $T$ is mostly fixed.

\begin{figure}
\centerline{\epsfig{file=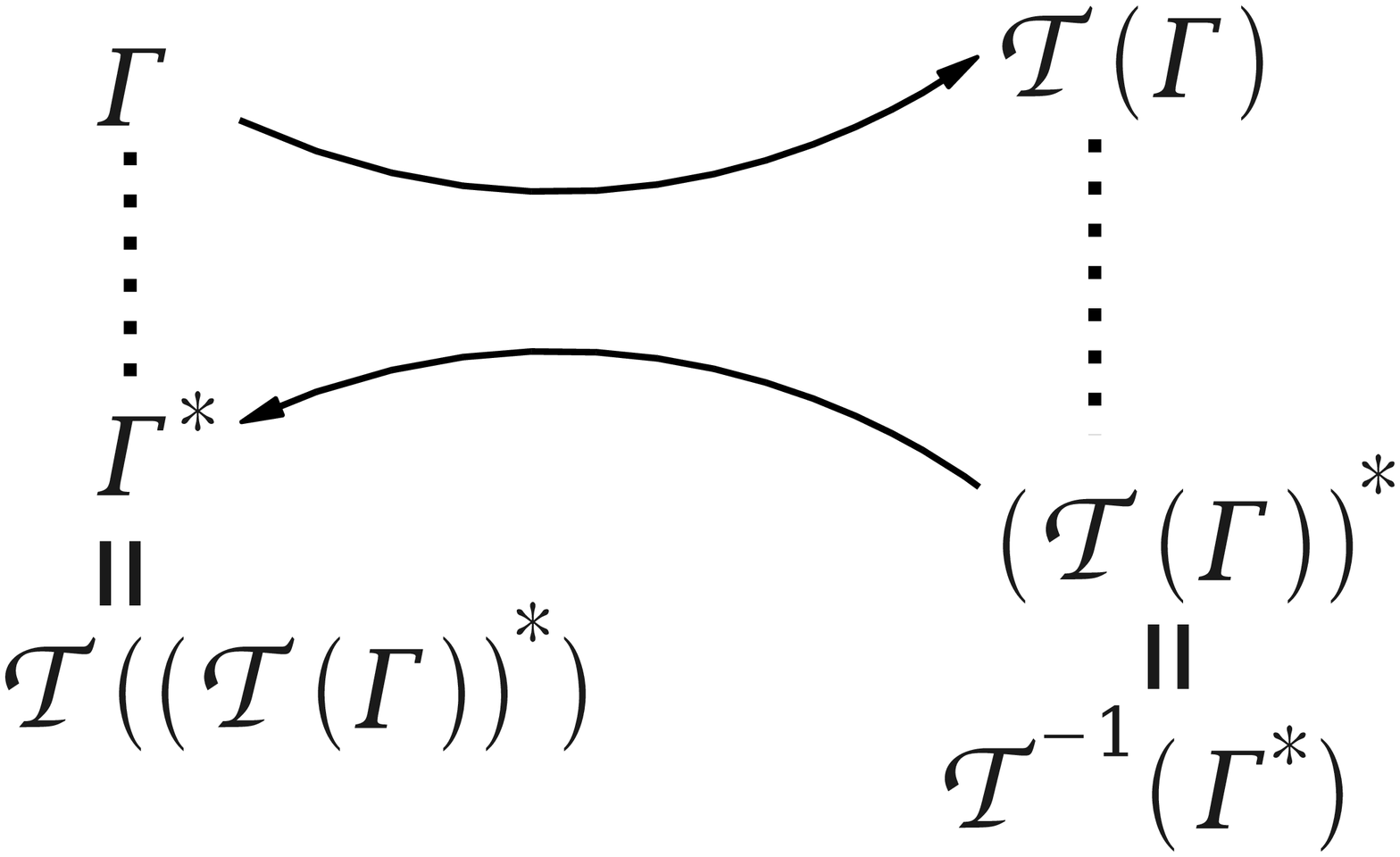,width=7cm}}
\caption[dummy]{
Schematics for the trajectory $\Gamma\rightsquigarrow\Te(\Gamma)$ 
and its time reversed trajectory $\Tei(\Gammas)\rightsquigarrow\Gammas$.
The state $\Gammas$ is the time reversal of the state $\Gamma$.
}
\label{f:path}
\end{figure}

The time evolution map $\Te$ may be that generated by the static Hamiltonian \eqref{e:Hstat}, but may be much more general (see section~\ref{s:examples} for examples).
All that we require are that $\Te$ preserves the Lebesgue measure of the total phase space $\calS$
 (which is the statement of the Liouville theorem), and that it satisfies the time reversal symmetry
\begin{equation}
(\Te(\Gamma))^*=\Tei(\Gammas),
\label{e:TRS}
\end{equation}
for any $\Gamma\in\calS$,
where $\Tei$ is the inverse time evolution, i.e., the inverse function of $\Te$.
See Fig.\ref{f:path}.
One has the symmetry \eqref{e:TRS} in a general Newtonian dynamics without a magnetic field.

For an arbitrary function $f(\cdot)$ on $\calS$, we define its time reversal $f^\dagger(\cdot)$ by
\begin{equation}
f^\dagger(\Gamma):=f(\Tei(\Gammas)),
\label{e:fbar}
\end{equation}
for any $\Gamma\in\calS$.
Intuitively speaking, $f^\dagger(\cdot)$ is basically the same thing as $f(\cdot)$, but represented as a function of the ``final state'' at $t=T$.

Since we are considering a deterministic mechanical system, the initial state $\Gamma$ determines the whole trajectory  from $t=0$ to $t=T$.
Some function $f(\Gamma)$ on $\calS$ should better be interpreted as  a function of the whole trajectory $\Gamma\rightsquigarrow\Te(\Gamma)$ rather than the initial state $\Gamma$.
(A notable example of such a function is the entropy production defined in \eqref{e:SG}.)
In such a case, the  time reversal $f^\dagger(\Gamma)$ is interpreted as a 
function of the time reversed trajectory $\Tei(\Gammas)\rightsquigarrow\Gammas$.

\subsection{Initial distribution and the steady state}
Let $\beta_1,\beta_2,\ldots,\beta_n$ be the inverse temperatures of the heat baths, which we shall fix.
Let us denote them collectively as  $\bsbeta=(\beta_1,\beta_2,\ldots,\beta_n)$.
For any given $\nu\in\Ss$, we define the initial distribution as
\begin{equation}
P_{\nu}(\Gamma)=\delta(\Gs-\nu)\,\frac{1}{Z_{\nu}(\bsbeta)}\,\exp\Bigl[-\sum_{i=1}^n\beta_i \,\{\Hb_i(\Gamma_i)+\Hc_i(\Gamma_i,\Gs)\}\Bigr],
\label{e:Pgamma}
\end{equation}
where the system is fixed at the given state $\nu$, and the heat baths are in the corresponding equilibrium with  inverse temperatures $\bsbeta$.
We do not make the $\bsbeta$ dependence of $P_{\nu}(\Gamma)$ explicit.
In \eqref{e:Pgamma}, the partition function $Z_{\nu}(\bsbeta)$
 is determined by requiring that $\int \dG P_{\nu}(\Gamma)=1$. 
From \eqref{e:Pgamma}, we see that
\begin{align}
Z_{\nu}(\bsbeta)&=\int \dG\,\delta(\Gs-\nu)\, \exp\Bigl[-\sum_{i=1}^n\beta_i\,\{\Hb_i(\Gamma_i)+\Hc_i(\Gamma_i,\Gs)\}\Bigr]
\ret
&=\prod_{i=1}^n\int \dG_i \,\exp[-\beta_i\,\{\Hb_i(\Gamma_i)+\Hc_i(\Gamma_i,\nu)\}].
\label{e:Phii}
\end{align}
Thus $Z_{\nu}(\bsbeta)$  is a product of the equilibrium partition functions with the ``boundary condition'' $\nu$.
Let
\begin{equation}
\tilde{Z}(\bsbeta):=\prod_{i=1}^n\int \dG_i \,\exp[-\beta_i\,\Hb_i(\Gamma_i)],
\label{e:Ztild}
\end{equation}
be the similar (but $\nu$ independent) partition function without coupling terms.
The ``free energy'' for the coupling between the system and the baths defined as
\begin{equation}
\phic({\nu})
:=
-\log\frac{Z_{\nu}(\bsbeta)}{\tilde{Z}(\bsbeta)},
\label{e:Zgf}
\end{equation}
appears in our representation.
Let us stress that
$\phic({\nu})$ is a combination of equilibrium quantities with different temperatures, and is in principle computable.
In the weak coupling limit, which is standard in the literature, one neglects the effect of the coupling $\Hc_i(\Gamma_i,\nu)$ (except for that needed to get steady states).
In this limit one can simply set  $\phic({\nu})=0$.
%

Let us assume that the state at $t=0$ is drawn from the distribution \eqref{e:Pgamma},
 and consider the state of the system at $t=T$. 
By definition the probability density of finding the system at state $\gamma\in\Ss$ is
\begin{equation}
\rho_{\nu}(\gamma)=\int \dG\,P_{\nu}(\Gamma)\,\delta(\ofs{\Te(\Gamma)}-\gamma).
\label{e:rhogg}
\end{equation}

We assume that the heat baths are so large that their states won't  change essentially for quite a long time (see the remark below).
This means that each heat bath essentially remains in the equilibrium
 with the inverse temperature specified in the initial state \eqref{e:Pgamma}.
Then there exists a range of time which is short enough for the heat baths but long enough for the system.
Within such a time scale, the system is expected to settle to a unique nonequilibrium steady state
 which is independent of the initial state $\nu$, and is characterized by the inverse temperatures
 $\bsbeta$ of the heat baths as well as other nonequilibrium conditions (see the examples below).

Suppose that $T$ is chosen from this range.
We can then reasonably assume that
\begin{equation}
\rho_{\nu}(\gamma)=\rhoss(\gamma),
\label{e:rhoss}
\end{equation}
where $\rhoss(\gamma)$ denotes the probability distribution in the unique nonequilibrium steady state.

\bigskip\noindent
{\bf Remark}:
It is possible to define an artificial heat bath which maintains equilibrium for an arbitrarily long time.
The bath consists of classical particles confined in a three dimensional box defined by $0\le x,y\le\ell$ and $0\le z\le L$.
The particles do not interact with each other and are reflected elastically by the walls.
The face with $z=0$ is attached to the system, and the particles in the system and those in the bath interact through short range repulsive interaction.

In the initial state \eqref{e:Pgamma}, the particles in the bath are uniformly distributed in the box (except near $z=0$), and their velocities are exactly distributed according to the Maxwell-Boltzmann distribution.
As the whole system evolves, the Maxwell-Boltzmann distribution may be lost by the interaction between the system and the bath.
But those ``nonequilibrium particles'' simply fly away to the positive $z$ direction, and won't come back until it is reflected back by the wall at $z=L$.
This means that the bath looks exactly as in equilibrium from the system for a finite amount of time.
By making $L$ large with the density fixed, we can realize a bath which is effectively in equilibrium for an arbitrarily long time.

We note that this bath, which lacks relaxation process, does not provide a  model of realistic large baths.
But it shows that our assumption is satisfied in at least one example.

\subsection{Conditioned average}
For an arbitrary function $f(\cdot)$ on $\calS$, and any $\nu,\gamma\in\Ss$,
we define the conditioned average
\begin{equation}
\sbkt{f}_{\nu,\gamma}:=
\frac{\int \dG\,f(\Gamma)\,P_{\nu}(\Gamma)\,\delta(\ofs{\Te(\Gamma)}-\gamma)}{\rho_{\nu}(\gamma)}.
\label{e:fgg}
\end{equation}
Note that this may be interpreted as the average over all the histories in which the system is initially at $\nu$ and finally at $\gamma$.
To consider the average in which both the initial and the final conditions are specified was an essential idea in \cite{KNPRL}, on which the present work is also based.

In addition to the conditioned average \eqref{e:fgg}, we define two kinds of partially conditioned averages.
One is
\begin{equation}
\sbkt{f}_{\nu,\sss}:=
\int \dd\gamma\,\rhoss(\gamma)\,\sbkt{f}_{\nu,\gamma}=
\int \dG\,f(\Gamma)\,P_{\nu}(\Gamma)\,
\label{e:fgs}
\end{equation}
which is a natural average when the initial distribution $P_{\nu}(\Gamma)$ is specified.
By taking into account the assumption that the unique steady state is attained at $t=T$,
 we have written the final state as ``$\sss$'' (which stands for steady state).
The other average treats the opposite situation, where the system starts from the steady state and ends precisely at the specified state $\gamma\in\Gs$.
It is defined by averaging $\sbkt{f}_{\nu,\gamma}$ over the initial state as
\begin{equation}
\sbkt{f}_{\sss,\gamma}:=
\int \dd\nu\,\rhoss(\nu)\,\sbkt{f}_{\nu,\gamma}.
\label{e:fsg}
\end{equation}

\subsection{Examples}
\label{s:examples}
Let us describe typical examples to which our general theory apply.

In what follows, we mean by ``equilibrium case'' the situation where $\beta_i=\betaeq$ for all $i$ with some $\betaeq>0$,
 and the time evolution $\Te$ is the pure Hamiltonian time evolution determined by the static Hamiltonian \eqref{e:Hstat}.

\begin{figure}
\centerline{\epsfig{file=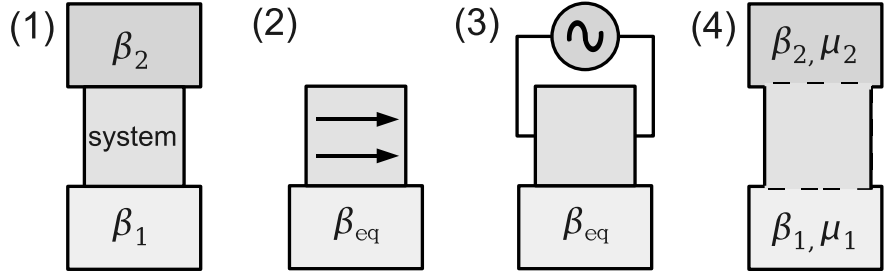,width=12cm}}
\caption[dummy]{
Four typical examples to which our general theory apply.
(1)~Heat conducting system described by a pure Hamiltonian mechanics.
(2)~Newtonian system with a constant driving force.
(3)~Hamiltonian system with an oscillating external field.
(4)~System with particle flow (and heat conduction) maintained by two reservoirs with different chemical potentials.
}
\label{f:123}
\end{figure}

We shall consider models in which certain nonequilibrium features are added to this  ``equilibrium case.''
See Fig.~\ref{f:123}.
There can be many examples that fit into our general scheme, but let us discuss three typical cases.
In all the examples (and in general), we denote by $\epsilon>0$ a dimensionless quantity
that measure the ``degree of nonequilibrium''.
The equilibrium case corresponds to $\epsilon=0$.

\bigno
{\em 1.~Heat conduction:}
\par\noindent
We let the number $n$ of the heat baths to be more than one, and assume that $\beta_1,\ldots,\beta_n$ are different.
The time evolution $\Te$ is the pure Hamiltonian time evolution determined by the static Hamiltonian \eqref{e:Hstat}.
Then there should be a nonvanishing heat current in the steady state.
We take  $\epsilon$ to be proportional to the heat current.

\bigno
{\em 2.~Driven system (DC field):}
\par\noindent
We let $n=1$, and $\beta_1=\betaeq$.
Suppose that the particles in the system are confined in a box with periodic boundary conditions in one direction, say the $x$-direction.
We assume that particles feel external driving forces (proportional to $\epsilon$) in the $x$-direction.
One may imagine that there is an ``electric field'' in the $x$-direction and exerts the electrostatic force on  charged particles.
Since such an external drive cannot be modeled by a Hamiltonian dynamics (when we use periodic boundary conditions), we consider a Newtonian dynamics in which the force on a particle is the sum of the external driving force and the force determined by the static Hamiltonian \eqref{e:Hstat}.
It is crucial that the Liouville theorem and the time reversal symmetry \eqref{e:TRS} still hold for this dynamics.

\bigno
{\em 3.~Driven system (AC field):}
\par\noindent
We again  let $n=1$, and $\beta_1=\betaeq$.
We consider the situation where particles are acted by a force which varies periodically in time.
A typical example is a system under an oscillating electric field.
After a sufficiently long time, the system is expected to settle into a periodically varying state with the same period as the external force.
By looking at the state when external force has a fixed phase, one effectively observes a ``steady'' state.
We wish to focus on such a sate (and call it a steady state with a slight abuse of the terminology).

The time evolution $\Te$ is determined by a time dependent Hamiltonian
\begin{equation}
\Hstat(\Gamma)+\epsilon\,V_t(\Gs),
\label{e:H+V}
\end{equation}
where $V_t(\Gs)$ is a periodically changing potential.
The dynamics satisfies the Liouville theorem.
In order to guarantee the time reversal symmetry \eqref{e:TRS},
 we further assume that the potential satisfies $V_t(\Gs)=V_{T-t}(\Gs)$.

\bigskip
Clearly one can consider models in which these three nonequilibrium factors (and other possible factors) are combined.
But we believe that to have these three typical examples in mind will be helpful in understanding the general representation that we shall derive.

In section~\ref{s:paricleflow}, we  treat a situation in which the system and the baths (reservoirs) exchange particles.
Then one can realize

\bigno
{\em 4.~Particle flow maintained by a difference in chemical potentials:}
By considering a system attached to multiple reservoirs with different chemical potentials (and possibly with different temperatures) and allowing the system and the reservoirs to exchange particles, one can realize a nonequilibrium steady state with a constant flow of particles (and possibly with a heat conduction).
See section~\ref{s:paricleflow} for details.

\section{Representations of $\rhoss(\gamma)$}
\label{s:Representaion}

In the present section, we  describe the representations \eqref{e:main1} and \eqref{e:main3} of the steady state distribution.
For this purpose we define relevant quantities carefully in section~\ref{s:EP}.
We also compare the present representation with the existing similar results in section~\ref{s:comp}.

\subsection{Entropy production and excess entropy production}
\label{s:EP}
For $\Gamma\in\calS$, let us define
\begin{equation}
\Theta(\Gamma):=\sum_{i=1}^n\beta_i\{H_i(\Te(\Gamma))-H_i(\Gamma)\},
\label{e:SG}
\end{equation}
where $H_i(\Gamma)$ is the total Hamiltonian for the $i$-th bath defined in \eqref{e:Hidef}.
Note that  $H_i(\Te(\Gamma))-H_i(\Gamma)$ can be interpreted as the total heat
 that has flown from the system to the $i$-th heat bath\footnote{
In this definition of heat, we interpreted the energy from the coupling Hamiltonian $\Hc_i(\Gamma_i,\Gs)$ as a part of the energy of the bath.
In fact we do not have a convincing physical argument for justifying this particular choice.
We use this definition only because we can derive the main representation by using it.
Of course this delicate issue becomes irrelevant in the weak coupling limit.
} during the time interval from $t=0$ to $t=T$.
Thus $\Theta(\Gamma)$ defined in \eqref{e:SG} is the total entropy production due to the transfer of heat into the $n$ heat baths.
Although $\Theta(\Gamma)$ is  defined here as a function of the initial state $\Gamma\in\calS$,
 it may be more naturally regarded as a function of the whole trajectory from $t=0$ to $t=T$.
In fact by denoting $j_i(t;\Gamma)$ the heat current that flows into the $i$-th bath at time $t$ in the trajectory $\Gamma\rightsquigarrow\Te(\Gamma)$, one has
\begin{equation}
\Theta(\Gamma)=\sum_{i=1}^n\beta_i\int_0^T\dt\,j_i(t;\Gamma)=\int_0^T\dt\,\sigma(t;\Gamma),
\label{e:Sint}
\end{equation}
where 
\begin{equation}
\sigma(t;\Gamma):=\sum_{i=1}^n\beta_i\,j_i(t;\Gamma)
\label{e:st}
\end{equation}
is the entropy production rate.

In a nonequilibrium steady state, there should be a constant flow of heat into (or from)
 heat baths resulting in positive constant entropy production rate.
Therefore the average like $\sbkt{\Theta}_{\nu,\gamma}$ should grow linearly with time $T$.
Let us define the entropy production rate $\barsigma$ as
\begin{equation}
\barsigma:=\lim_{T\to\infty}\frac{1}{T}\sbkt{\Theta}_{\cdot,\cdot}
\label{e:sigmadef}
\end{equation}
where the average may be any of the three kinds   \eqref{e:fgg}, \eqref{e:fgs}, \eqref{e:fsg} that we have defined since the long time behavior should not depend on the initial and the final conditions.
Here $T\to\infty$  means ``make $T$ large  in the range which is not too large for the heat baths.''

Then by subtracting the steady contribution from the total entropy production, we define the excess entropy production as
\begin{equation}
\Sex(\Gamma):=\Theta(\Gamma)-\barsigma\,T=\int_0^T\dt\,\{\sigma(t;\Gamma)-\barsigma\}.
\label{e:Sex}
\end{equation}
We recall that the similar excess quantities plays a fundamental role
 in the phenomenological approach to nonequilibrium steady states \cite{OP,HS}.

\subsection{The second order representations}

Now we shall state our major result.
Let us write the steady state distribution as
\begin{equation}
\rhoss(\gamma)=\exp[-\pss(\gamma)],
\label{e:rhosspss}
\end{equation}
where $\pss(\gamma)$ is the ``effective Hamiltonian'' which plays a role of the quantity $\betaeq\Hs(\gamma)$ in  equilibrium.

The most important result of the present paper is the concise representation (obtained in \cite{KNPRL}) of the steady state distribution
\begin{equation}
\pss(\gamma)=\phizero+\phic({\gamma})+
\frac{1}{2}\{\sbkt{\Sex}_{\gammas,\sss}-\sbkt{\Sex}_{\sss,\gamma}\}
+O(\epsilon^3),
\label{e:main1}
\end{equation}
where $\phizero$ is a normalization constant\footnote{
In \cite{KNSTshort}, we found that the normalization constant plays the role of the entropy of the nonequilibrium state.
} independent of $\gamma$ (but of course dependent on $\bsbeta$ and other nonequilibrium parameters).
The ``free energy'' of the coupling $\phic({\gamma})$ defined in \eqref{e:Zgf} is an equilibrium quantity which becomes negligible in the  weak coupling limit.
Thus 
the only nonequilibrium quantities contained in the right-hand side of \eqref{e:main1}
 are the expectation values of the excess entropy production $\Sex$.
Since $\Sex$ may be regarded as a quantity of first order in $\epsilon$,
 one may naturally expect that the representation \eqref{e:main1} is correct up to the first order in $\epsilon$.
Rather surprisingly, we shall show that (as indicated in the error term
 in \eqref{e:main1}) this representation is correct to the second order in the degree of nonequilibrium $\epsilon$.

We indeed have an expression to the full order in $\epsilon$ as in \eqref{e:finalfull}.
But the above second order expression seems most useful and suggestive.

Let us examine the two conditioned expectation values that appear in \eqref{e:main1}.
Recalling the expression \eqref{e:Sex} of the excess entropy production, the first expectation value is written as
\begin{equation}
\sbkt{\Sex}_{\gammas,\sss}=\int_0^T\dt\,\sbkt{\sigma(t)-\barsigma}_{\gammas,\sss}.
\label{e:Sexint2}
\end{equation}
Since the average of $\sigma(t;\Gamma)$ in the steady state is $\barsigma$, the integrand is vanishing except for small values of $t$ where the system is forced to be out of the steady state by the imposed initial condition.
Thus the integrand in \eqref{e:Sexint2} is non-negligible only for $t\le\taus$, where $\taus$ is a constant which is sufficiently larger than the relaxation time of the system.
Likewise the other average $\sbkt{\Sex}_{\sss,\gamma}$ is essentially determined from an integral from $T-\taus$ to $T$.
We thus find that the representation \eqref{e:main1} converges rapidly when $T$ is increased beyond the relaxation time.

The representation \eqref{e:main1} has a very interesting form which contains the difference between the two conditioned averages of the excess entropy production.
Since it turns out that $\sbkt{\Sex}_{\gammas,\sss}=-\sbkt{\Sex}_{\sss,\gamma}+O(\epsilon^2)$, the two terms roughly have comparable contributions.
But to get the result which is precise to $O(\epsilon^2)$, it is necessary to consider the difference of the two averages.

\subsection{Relation to the canonical distribution}
\label{s:canonical}
By fixing a reference equilibrium inverse temperature $\betaeq$, and using the energy conservation, we can rewrite the representation \eqref{e:main1} in a different form, in which the contributions from the canonical distribution becomes clearer.

Let us define
\begin{equation}
\Phi(\Gamma):=\Theta(\Gamma)+\betaeq\{\Hs(\ofs{\Te(\Gamma)})-\Hs(\Gs)\},
\label{e:Sneq}
\end{equation}
which we may call the ``nonequilibrium part'' of the entropy production.
Here $\betaeq$ is the reference inverse temperature, which may be chosen rather arbitrarily.

By recalling the definitions of the static Hamiltonian \eqref{e:Hstat} and the entropy production \eqref{e:SG}, we see that
\begin{equation}
\Phi(\Gamma)=\betaeq W(\Gamma)+\sum_{i=1}^n\Di\beta_i\{H_i(\Te(\Gamma))-H_i(\Gamma)\},
\label{e:Sneq2}
\end{equation}
where $\Di\beta_i=\beta_i-\betaeq$ and  $W(\Gamma)=\Hstat(\Te(\Gamma))-\Hstat(\Gamma)$
 is the total work that was done (say, by the external force) to the whole system from $t=0$ to $t=T$.
Note that in the example 1 of heat conduction, we have $W(\Gamma)=0$ because of the energy conservation.
On the other hand, in the example 2 and 3, we have $\Di\beta_i=0$ and the second term of \eqref{e:Sneq2} is vanishing.

Since $\Theta(\Gamma)$ and $\Phi(\Gamma)$ differs only by a quantity which depends on the initial and final states of the system,  the averages $\sbkt{\Phi}_{\gammas,\sss}$,  $\sbkt{\Phi}_{\sss,\gamma}$ grow linearly as $\barsigma\,T$ with the same $\barsigma$ as in \eqref{e:sigmadef}.
We therefore define the corresponding excess quantity as  $\Phi_\mathrm{ex}(\Gamma):=\Phi(\Gamma)-\barsigma\,T$.
Then the representation \eqref{e:main1} is rewritten in the form
\begin{equation}
\pss(\gamma)=\varphi^{(1)}+\phic({\gamma})+\betaeq\,\Hs(\gamma)
+\frac{1}{2}\{\sbkt{\Phi_\mathrm{ex}}_{\gammas,\sss}-\sbkt{\Phi_\mathrm{ex}}_{\sss,\gamma}\}
+O(\epsilon^3),
\label{e:main3}
\end{equation}
where the new constant is defined by $\varphi^{(1)}:=\phizero-\betaeq\int d\gamma\rhoss(\gamma)\Hs(\gamma)$.


\subsection{Comparison with known formulas}
\label{s:comp}
We shall here compare our representation \eqref{e:main1} and \eqref{e:main3} with two of the well-known representations of the steady  distribution.

First is the linear response theory.
Indeed it is automatic to get a representation correct up to $O(\epsilon)$ if we have a representation which is correct  to $O(\epsilon^2)$.
By starting from the representation \eqref{e:main3}, and only taking the lowest order contributions, one arrives at
\begin{equation}
\pss(\gamma)=\varphi^{(1)}+\phic({\gamma})+\betaeq\,\Hs(\gamma)+
\frac{1}{2}\{\sbkt{\Phi}_{\gammas,\eqs}^{\epsilon=0}-\sbkt{\Phi}_{\eqs,\gamma}^{\epsilon=0}\}
+O(\epsilon^2).
\label{e:LRpre}
\end{equation}
Here $\sbkt{\cdots}^{\epsilon=0}_{\cdot,\cdot}$ denotes the conditioned averages as in \eqref{e:fgs} and \eqref{e:fsg}, defined in the corresponding ``equilibrium case'' with $\epsilon=0$.
We also replace the steady state in the initial or final conditions by the equilibrium state (as is indicated by the subscript ``eq'').
The averaged quantity $\Phi$ in \eqref{e:LRpre}, on the other had, is a genuine nonequilibrium quantity \eqref{e:Sneq2}.
By using the time reversal symmetry \eqref{e:f=f}, which implies 
$\sbkt{\Phi}_{\gammas,\eqs}^{\epsilon=0}=-\sbkt{\Phi}_{\eqs,\gamma}^{\epsilon=0}$, 
the expression \eqref{e:LRpre} can be simplified as
\begin{equation}
\pss(\gamma)=\varphi^{(1)}+\phic({\gamma})+\betaeq\,\Hs(\gamma)
-
\sbkt{\Phi}_{\eqs,\gamma}^{\epsilon=0}
+O(\epsilon^2).
\label{e:LR}
\end{equation}

The expression \eqref{e:LR} is the standard result of the linear response theory, and may be used to derive well-known useful expressions of, say, transport coefficients.
We wish to stress that our representation \eqref{e:main3} is as concise as the linear response relation \eqref{e:LR} but properly takes into account nonlinear effects.

Next we shall focus on exact expressions for the steady state distribution.
By using the same notations as before, one can (rather formally) show that the steady state distribution is written as
\begin{equation}
\rhoss(\gamma)=\mathrm{const.}\,Z_{\gamma}(\bsbeta)\,\sbkt{e^{-\Theta}}_{\gammas,\gamma_0},
\label{e:exact}
\end{equation}
where $\gamma_0\in\Ss$ is an arbitrary fixed reference state.
See the end of section~\ref{s:repder}.
This is one of the well-known exact expressions for steady state distribution
 which were discussed by many authors including Zubarev, McLennan, and Kawasaki and Gunton
\cite{McLennan90,Zubarev74,KawasakiGunton73}
.

Although \eqref{e:exact} somehow resembles our \eqref{e:main1}, there are indeed marked differences.
First of all, \eqref{e:exact} is an expression for the distribution itself and hardly provides information about the ``effective Hamiltonian'' $\pss(\gamma)=-\log\rhoss(\gamma)$.
Moreover it is crucial that the bare entropy production $\Theta$ (rather than the excess entropy production $\Sex$) appears in \eqref{e:exact}.
Since the quantity $\Theta$ typically grows as $\barsigma\,T$ in the steady state, the $T\to\infty$ limit of \eqref{e:exact} is extremely delicate.
It is expected that the quantity $e^{-\Theta}$ exhibits wild fluctuation, and a miraculous cancellation leads to a result which is independent of $T$.

On the other hand,  our expression \eqref{e:main1} has a nicely controlled large $T$ behavior.
What we have in \eqref{e:main1}  are essentially short time integrals of the excess entropy production.
This is indeed true for our higher order expressions \eqref{e:finalfull}.

To conclude, although the exact but formal expression \eqref{e:exact} and our representation \eqref{e:main1} apparently look similar, their natures are drastically different.

\section{Derivation}
\label{s:Derivation}
Let us discuss the derivation of the results in detail.

In section~\ref{s:BI}, we derive the basic identity \eqref{e:ZrfZrf}, which represents the time-reversal symmetry in the present system.
The derivation is a straightforward application of the standard idea repeatedly used, for example, in \cite{Crooks,J,MN}.

In sections~\ref{s:repder} and \ref{s:OE}, we derive the representation \eqref{e:main1}.
The basic idea is essentially the same as that in \cite{KNPRL}, but we have refined the derivation to make it as efficient and transparent as possible.
We note that although our derivation is quite sensible in physicists' standard, it is not (yet) a mathematical proof.
We believe that some new ideas are required to construct a true proof.

\subsection{Basic Identity}
\label{s:BI}
We first show the identity
\begin{equation}
Z_{\nu}\,\rho_{\nu}(\gamma)\,\sbkt{f\,e^{-\Theta/2}}_{\nu,\gamma}
=
Z_{\gammas}\,\rho_{\gammas}(\nus)\,\sbkt{f^\dagger\,e^{-\Theta/2}}_{\gammas,\nus},
\label{e:ZrfZrf}
\end{equation}
which is valid for an arbitrary function $f(\cdot)$ on $\calS$.
Here and in what follows, we drop the $\bsbeta$ dependence of $Z_\nu$.
We note that this identity is a formal consequence of mechanics (and the choice of initial distribution), and is independent of the assumption \eqref{e:rhoss} about the approach to steady state.
Let us also note that by dividing \eqref{e:ZrfZrf} by the same equation with $f(\cdot)=1$, we get an interesting identity
\begin{equation}
\frac{\sbkt{f\,e^{-\Theta/2}}_{\nu,\gamma}}{\sbkt{e^{-\Theta/2}}_{\nu,\gamma}}
=\frac{\sbkt{f^\dagger\,e^{-\Theta/2}}_{\gammas,\nus}}{\sbkt{e^{-\Theta/2}}_{\gammas,\nus}},
\label{e:feoe}
\end{equation}
which reveals a beautiful time-reversal symmetry in a system far from equilibrium.

Let us show \eqref{e:ZrfZrf}.
From the definitions \eqref{e:Pgamma} and \eqref{e:fgg}, we find
\begin{equation}
Z_{\nu}\,\rho_{\nu}(\gamma)\,\sbkt{f}_{\nu,\gamma}
=
\int \dG\,f(\Gamma)\,\exp\Bigl[-\sum_{i=1}^n\beta_i\,
H_i(\Gamma)\Bigr]\,\delta(\Gs-\nu)\,\delta(\ofs{\Te(\Gamma)}-\gamma).
\label{e:Zrf00}
\end{equation}
Then by using this and the definition \eqref{e:SG} of the entropy production,
the left-hand side of \eqref{e:ZrfZrf} becomes
\begin{align}
&Z_{\nu}\,\rho_{\nu}(\gamma)\,\sbkt{f\,e^{-\Theta/2}}_{\nu,\gamma}
\ret
&=
\int \dG\,f(\Gamma)\,\exp\Bigl[-\frac{1}{2}\sum_{i=1}^n\beta_i
\{H_i(\Gamma)+H_i(\Te(\Gamma))\}\Bigr]\,\delta(\Gs-\nu)\,\delta(\ofs{\Te(\Gamma)}-\gamma)
\notag
\intertext{Here we shall make a change of variable according to $\Upsilon=\Te(\Gamma)$.
Since the measure preserving nature of $\Te$ ensures $\dG=\dd\Upsilon$, we have
}
&=
\int \dd\Upsilon\,f(\Tei(\Upsilon))\,\exp\Bigl[-\frac{1}{2}\sum_{i=1}^n\beta_i
\{H_i(\Tei(\Upsilon))+H_i(\Upsilon)\}\Bigr]\,\delta(\ofs{\Tei(\Upsilon)}-\nu)\,\delta(\Upsilon_\mathrm{s}-\gamma)
\notag
\intertext{By noting that $\Tei(\Upsilon)=(\Te(\Upsilons))^*$ (as in \eqref{e:TRS}),
 $f(\Tei(\Upsilon))=f^\dagger(\Upsilons)$ (as in \eqref{e:fbar}),
 $\dd\Upsilon=\dd\Upsilons$, and $H_i(\Gamma)=H_i(\Gammas)$, we can rewrite the above as
}
&=
\int \dd\Upsilons\,f^\dagger(\Upsilons)\,\exp\Bigl[-\frac{1}{2}\sum_{i=1}^n\beta_i
\{H_i(\Te(\Upsilons))+H_i(\Upsilons)\}\Bigr]\,\delta(\ofs{\Te(\Upsilons)}-\nus)\,\delta(\ofs{\Upsilons}-\gammas)
\notag
\intertext{By  rewriting $\Upsilons$ as $\Gamma$, we have}
&=
\int \dG\,f^\dagger(\Gamma)\,\exp\Bigl[-\frac{1}{2}\sum_{i=1}^n\beta_i
\{H_i(\Gamma)+H_i(\Te(\Gamma))\}\Bigr]\,\delta(\Gs-\gammas)\,\delta(\ofs{\Te(\Gamma)}-\nus)
\notag
\intertext{By comparing this with the second line of the present equation, we find that}
&=Z_{\gammas}\,\rho_{\gammas}(\nus)\,\sbkt{f^\dagger\,e^{-\Theta/2}}_{\gammas,\nus}.
\label{e:ZrfZrf2}
\end{align}

\subsection{Derivation of the representation}
\label{s:repder}
By setting $f(\cdot)=1$ in \eqref{e:ZrfZrf}, and noting that  $Z_{\gamma}=Z_{\gammas}$ and $Z_{\nu}=Z_{\nu^*}$ because the Hamiltonians are symmetric under time reversal, we see that
\begin{equation}
\frac{\rho_{\gammas}(\nus)}{\rho_{\nu}(\gamma)}=
\frac{Z_{\nus}}{Z_{\gamma}}\,
\frac{\sbkt{e^{-\Theta/2}}_{\nu,\gamma}}{\sbkt{e^{-\Theta/2}}_{\gammas,\nus}}.
\label{e:rrZZee}
\end{equation}
By writing $\rho_{\nu}(\gamma)=\exp[-\varphi_{\nu}(\gamma)]$, and recalling \eqref{e:Zgf}, we have
\begin{equation}
\varphi_{\nu}(\gamma)-\varphi_{\gammas}(\nus)
=\phic({\gamma})-\phic({\nus})
+\log\sbkt{e^{-\Theta/2}}_{\nu,\gamma}
-\log\sbkt{e^{-\Theta/2}}_{\gammas,\nus}.
\label{e:ppFFll}
\end{equation}

To proceed further we make use of the standard technique of cumulant expansion.
For any random variable $Y$ (associated with an average $\sbkt{\cdots}$) and a positive integer $k$,
 we define the $k$-th order cumulant of $Y$ by
\begin{equation}
\sbkt{Y^k}^\mathrm{c}:=\left.\frac{\partial^k}{\partial u^k}\log\sbkt{\exp[u\,Y]}\right|_{u=0}.
\label{e:Ykc}
\end{equation}
Then one has a formal Taylor expansion
\begin{equation}
\log{\sbkt{e^Y}}=\sum_{k=1}^\infty\frac{1}{k!}\sbkt{Y^k}^\mathrm{c}.
\label{e:eYexp}
\end{equation}
By using \eqref{e:Ykc}, one has 
\begin{equation}
\sbkt{Y}^\mathrm{c}=\sbkt{Y},\quad
\sbkt{Y^2}^\mathrm{c}=\sbkt{Y^2}-(\sbkt{Y})^2.
\label{e:cummulant}
\end{equation}
It is also useful to note that for any nonrandom $y_0$, \eqref{e:Ykc} implies
\begin{equation}
\sbkt{(Y-y_0)^k}^\mathrm{c}=\sbkt{Y^k}^\mathrm{c},
\label{e:Yy0}
\end{equation}
for $k=2,3,\ldots$ (but not for $k=1$).

By applying the formal expansion \eqref{e:eYexp} to \eqref{e:ppFFll}, we have
\begin{equation}
\varphi_{\nu}(\gamma)-\varphi_{\gammas}(\nus)
=\phic({\gamma})-\phic({\nus})
+\sum_{k=1}^\infty
\frac{(-1)^k}{2^k\,k!}
\{
\sbkt{\Theta^k}^\mathrm{c}_{\nu,\gamma}
-\sbkt{\Theta^k}^\mathrm{c}_{\gammas,\nus}\}.
\label{e:ppFFSS}
\end{equation}

Now let us assume that $T$ is large enough and nonequilibrium steady state is realized in the system.
Then we can replace the distribution $\varphi_{\nu}(\gamma)$ by $\pss(\gamma)$.

We see that the cumulant expansion in \eqref{e:ppFFSS} also simplifies in this limit.
Let us examine the terms with $k=1$.
By using the integral representation \eqref{e:Sint} of $\Theta(\Gamma)$, we have
\begin{align}
\sbkt{\Theta}_{\nu,\gamma}-\sbkt{\Theta}_{\gammas,\nus}&=
\int_0^T \dt\{\sbkt{\sigma(t)}_{\nu,\gamma}-\sbkt{\sigma(t)}_{\gammas,\nus}\}
\ret
&=
\int_0^T \dt\{\sbkt{\sigma(t)-\barsigma}_{\nu,\gamma}-\sbkt{\sigma(t)-\barsigma}_{\gammas,\nus}\},
\label{e:SSk1}
\end{align}
where $\barsigma$ is the entropy production rate \eqref{e:sigmadef}.
Since the average of $\sigma(t;\Gamma)$ equals $\barsigma$ in the steady state, the expectation values in the right-hand side of \eqref{e:SSk1} are nonvanishing only for  $t$ which are either very close to 0 or very close to $T$
(see the discussion below \eqref{e:Sexint2}).
This means that we can safely decompose the expectation value as
\begin{align}
\int_0^T \dt\, \sbkt{\sigma(t)-\barsigma}_{\nu,\gamma}&=
\int_0^T \dt\, \sbkt{\sigma(t)-\barsigma}_{\nu,\sss}+
\int_0^T \dt\, \sbkt{\sigma(t)-\barsigma}_{\sss,\gamma}
\ret
&=\sbkt{\Sex}_{\nu,\sss}+\sbkt{\Sex}_{\sss,\gamma}
\label{e:sssdecomp}
\end{align}
where the expectation values are defined in \eqref{e:fgs} and \eqref{e:fsg}.
By using the similar decomposition for the other expectation value, \eqref{e:SSk1} leads to
\begin{equation}
\sbkt{\Theta}_{\nu,\gamma}-\sbkt{\Theta}_{\gammas,\nus}
=\sbkt{\Sex}_{\nu,\sss}+\sbkt{\Sex}_{\sss,\gamma}-\sbkt{\Sex}_{\gammas,\sss}-\sbkt{\Sex}_{\sss,\nus}
\label{e:SSSSSS}
\end{equation}
where we used the definition \eqref{e:Sex} of the excess entropy production.

We will later show (see section~\ref{s:cummulant}) that analogous result
\begin{equation}
\sbkt{\Theta^k}^\mathrm{c}_{\nu,\gamma}
-\sbkt{\Theta^k}^\mathrm{c}_{\gammas,\nus}
=\sbkt{\Theta^k}^\mathrm{c}_{\nu,\sss}+\sbkt{\Theta^k}^\mathrm{c}_{\sss,\gamma}
-\sbkt{\Theta^k}^\mathrm{c}_{\gammas,\sss}-\sbkt{\Theta^k}^\mathrm{c}_{\sss,\nus}
\label{e:SkSkSkSk}
\end{equation}
holds for $k\ge2$.

By substituting \eqref{e:SSSSSS} and \eqref{e:SkSkSkSk} into \eqref{e:ppFFSS}, we get
\begin{align}
\pss(\gamma)-\pss(\nus)
&=\phic({\gamma})-\phic({\nus})
\ret
&+\sum_{k=1}^\infty
\frac{(-1)^k}{2^k\,k!}
\{
\sbkt{(\Sex)^k}^\mathrm{c}_{\nu,\sss}+\sbkt{(\Sex)^k}^\mathrm{c}_{\sss,\gamma}
-\sbkt{(\Sex)^k}^\mathrm{c}_{\gammas,\sss}-\sbkt{(\Sex)^k}^\mathrm{c}_{\sss,\nus}\},
\label{e:ppFFSS2}
\end{align}
where we noted that \eqref{e:Yy0} implies $\sbkt{\Theta^k}^\mathrm{c}=\sbkt{(\Sex)^k}^\mathrm{c}$ for $k\ge2$.
Note that \eqref{e:ppFFSS2} is a sum of quantities each of which depends either on $\gamma$ or $\nu$.
Since \eqref{e:ppFFSS2} is valid for any $\gamma,\nu\in\Ss$, we must have
\begin{equation}
\pss(\gamma)=\phizero
+\phic({\gamma})+\sum_{k=1}^\infty
\frac{(-1)^k}{2^k\,k!}
\{
\sbkt{(\Sex)^k}^\mathrm{c}_{\sss,\gamma}
-\sbkt{(\Sex)^k}^\mathrm{c}_{\gammas,\sss}\},
\label{e:finalfull}
\end{equation}
for any $\gamma\in\Ss$ with a constant $\phizero$.
This is our full-order representation for the ``effective Hamiltonian'' $\pss(\gamma)$ of the nonequilibrium steady state.
Below in section~\ref{s:cummulant}, we shall see that both $\sbkt{(\Sex)^k}^\mathrm{c}_{\sss,\gamma}$ and $\sbkt{(\Sex)^k}^\mathrm{c}_{\gammas,\sss}$ grow linearly with $T$ for $k\ge2$, but  the difference has a  well-behaved  large $T$ limit.
Let us also note that one can formally rewrite \eqref{e:finalfull} into a compact form as
\begin{equation}
\pss(\gamma)=\phizero+\phic({\gamma})
-\log\frac{\sbkt{e^{-\Sex/2}}_{\sss,\gamma}}{\sbkt{e^{-\Sex/2}}_{\gammas,\sss}}.
\label{e:finalfull2}
\end{equation}

Finally let us show \eqref{e:exact} for completeness.
By setting $f(\Gamma)=\exp[\Theta(\Gamma)/2]$ in \eqref{e:ZrfZrf}, and using \eqref{e:rhoss}, one gets
\begin{equation}
Z_{\nu}\,\rho_{\nu}(\gamma)\,\sbkt{1}_{\nu,\gamma}
=
Z_{\gamma^*}\,\rho_{\gammas}(\nus)\,\sbkt{e^{-\Theta}}_{\gammas,\nus}.
\label{e:ZrfZrf3}
\end{equation}
Then we fix $\nu$ to an arbitrary constant $\gamma_0$ to get \eqref{e:exact}.

\bigskip\noindent
{\bf Remark:}
The full order expression like \eqref{e:finalfull} is indeed not unique.
Take a real constant $\alpha$ with $0\le\alpha\le1$.
By setting $f=\exp[\{\alpha-(1/2)\}\,\Theta]$ in \eqref{e:ZrfZrf}, one gets
\begin{equation}
\frac{\rho_{\gammas}(\nus)}{\rho_{\nu}(\gamma)}=
\frac{Z_{\nus}}{Z_{\gamma}}\,
\frac{\sbkt{e^{-\alpha\,\Theta}}_{\nu,\gamma}}{\sbkt{e^{-(1-\alpha)\,\Theta}}_{\gammas,\nus}}.
\label{e:rrZZee2}
\end{equation}
By combining this with the similar relation obtained by replacing $\alpha$ with $1-\alpha$,
we have
\begin{equation}
\frac{\rho_{\gammas}(\nus)}{\rho_{\nu}(\gamma)}=
\frac{Z_{\nus}}{Z_{\gamma}}\,\sqrt{
\frac{\sbkt{e^{-\alpha\,\Theta}}_{\nu,\gamma}\,\,\sbkt{e^{-(1-\alpha)\,\Theta}}_{\nu,\gamma}}
{\sbkt{e^{-\alpha\,\Theta}}_{\gammas,\nus}\,\,\sbkt{e^{-(1-\alpha)\,\Theta}}_{\gammas,\nus}}
}.
\label{e:rrZZee3}
\end{equation}
This reduces to \eqref{e:rrZZee} if we set $\alpha=1/2$.

By using \eqref{e:rrZZee3} instead of \eqref{e:rrZZee}, we get an expression corresponding to \eqref{e:finalfull}.
It turns out that the term with $k=1$ is the same as in \eqref{e:finalfull}, but the terms with $k\ge2$ are different.
Note that the expression need not be unique since we are not performing a naive power series expansion.
We have preliminary numerical evidences (in the Langevin models) which suggest
 that the series converges most efficiently when we set $\alpha=1/2$ and use \eqref{e:finalfull}.

\subsection{Order estimate}
\label{s:OE}
Let us show that, as we have noted in \eqref{e:main1},  the representation \eqref{e:finalfull} truncated to  include only the $k=1$ terms gives a result which is precise to the second order in the ``order of nonequilibrium'' $\epsilon$.

For this purpose we first note a well-known time-reversal symmetry in the equilibrium case with $\epsilon=0$, where the inverse temperatures of all the heat baths are identical to $\betaeq$ and the time evolution $\Te$ is completely determined by the static Hamiltonian $\Hstat(\Gamma)$ of \eqref{e:Hstat}.
Then one has  $W(\Gamma)=\Hstat(\Te(\Gamma))-\Hstat(\Gamma)=0$ from the energy conservation, and $\Di\beta_i=\beta_i-\betaeq=0$.
Thus the nonequilibrium entropy production of \eqref{e:Sneq2} is vanishing.
This, with \eqref{e:Sneq}, implies that
$\Theta(\Gamma)=-\betaeq\{\Hs(\ofs{\Te(\Gamma)})-\Hs(\Gs)\}$.
Therefore $\Theta(\Gamma)$ is entirely determined by the initial and the final states of the system, and shows no fluctuation within the averages $\sbkt{\cdots}_{\nu,\gamma}$ or $\sbkt{\cdots}_{\gammas,\nus}$.
Thus in the identity \eqref{e:feoe}, $e^{-\Theta/2}$ simply cancel out and we have
\begin{equation}
\sbkt{f}^{\epsilon=0}_{\nu,\gamma}=\sbkt{f^\dagger}^{\epsilon=0}_{\gammas,\nus}
\label{e:f=f}
\end{equation}
for any function $f(\cdot)$.

Let us evaluate the $k=2$ terms of \eqref{e:ppFFSS}.
Note that  \eqref{e:Sneq} implies
$\Phi_\mathrm{ex}(\Gamma)=\Sex(\Gamma)+\betaeq\{\Hs(\ofs{\Te(\Gamma)})-\Hs(\Gs)\}$.
Since the term $\betaeq\{\Hs(\ofs{\Te(\Gamma)})-\Hs(\Gs)\}$ is a constant in the conditioned average \eqref{e:fgg}, we see from \eqref{e:Yy0} that the second order terms in \eqref{e:ppFFSS} are written as
 ($1/8$ times) $\sbkt{(\Phi_\mathrm{ex})^2}^\mathrm{c}_{\nu,\gamma}-\sbkt{(\Phi_\mathrm{ex})^2}^\mathrm{c}_{\gammas,\nus}$.
Expanding this quantity around the equilibrium, one finds
\begin{align}
\sbkt{(\Phi_\mathrm{ex})^2}^\mathrm{c}_{\nu,\gamma}
-
\sbkt{(\Phi_\mathrm{ex})^2}^\mathrm{c}_{\gammas,\nus}
&=
\sbkt{(\Phi_\mathrm{ex})^2}^{\mathrm{c},\epsilon=0}_{\nu,\gamma}
-
\sbkt{(\Phi_\mathrm{ex})^2}^{\mathrm{c},\epsilon=0}_{\gammas,\nus}
+O(\epsilon^3),\notag
\intertext{where the correction is $O(\epsilon^3)$ because $\Phi_\mathrm{ex}$ itself is a quantity of $O(\epsilon)$.
By using \eqref{e:Yy0}, we replace  $\Phi_\mathrm{ex}$  by $\Phi$ to write
}
&=
\sbkt{\Phi^2}^{\mathrm{c},\epsilon=0}_{\nu,\gamma}
-
\sbkt{\Phi^2}^{\mathrm{c},\epsilon=0}_{\gammas,\nus}
+O(\epsilon^3)
\label{e:Sneqexp}
\end{align}
But by using \eqref{e:f=f} and $\Phi^\dagger=-\Phi$, we see that the first and the second terms in the right-hand side of  \eqref{e:Sneqexp} cancel with each other.
Thus the $k=2$ terms have only $O(\epsilon^3)$ contribution.
Since the terms with $k\ge3$ obviously are $O(\epsilon^3)$, we have shown the desired claim.

\subsection{More on cumulants}
\label{s:cummulant}
Let us complete some estimates related to cumulant.

We begin by introducing a general definition and a new notation.
For $k$ random variables $X_1,X_2,\ldots,X_k$, we define their cumulant by
\begin{equation}
\sbkt{X_1;X_2;\cdots;X_k}:=\left.\frac{\partial}{\partial u_1}\frac{\partial}{\partial u_2}\cdots\frac{\partial}{\partial u_k}
\log\sbkt{\exp[\sum_{i=1}^ku_i\,X_i]}\right|_{u_1=u_2=\cdots=u_k=0}.
\label{e:X1X2}
\end{equation}
The previous definition \eqref{e:Ykc} is a special case of the present one.
It is easily seen that one has
\begin{equation}
\sbkt{Y^k}^\mathrm{c}=\sbkt{\,\underbrace{Y;Y;\cdots;Y}_{k}\,}.
\label{e:YkYY}
\end{equation}
One easily find that the relation
\begin{equation}
\sbkt{X_1+X_1';X_2;\cdots;X_k}=\sbkt{X_1;X_2;\cdots;X_k}+\sbkt{X_1';X_2;\cdots;X_k}
\end{equation}
holds and that the cumulant is invariant under permutation of the order of $X_i$'s.

Using the integral representation \eqref{e:Sint}, we can write the cumulant of interest as
\begin{equation}
\sbkt{\Theta^k}^\mathrm{c}_{\nu,\gamma}
=\int_0^T \dt_1\int_0^T\dt_2\cdots\int_0^T\dt_k
\,\sbkt{\sigma(t_1);\sigma(t_2);\cdots;\sigma(t_k)}_{\nu,\gamma}
\label{e:Skint}
\end{equation}

It is reasonable to expect that the time evolution within the system resembles that of a stochastic process with a finite relaxation time.
More precisely we assume that there is a finite relaxation time\footnote{
It is well-known that most (stochastic) systems of particle exhibit power law decay of correlations, known as ``long-time tail'', in the infinite volume limit.
If a system is finite, however, one generally has an exponential decay, where the relaxation time $\tau$ is typically very large and diverges in the infinite volume limit.
} $\tau$ and finite constants $C_2,C_3,\ldots$, and one has
\begin{equation}
\abs{\sbkt{\sigma(t_1);\sigma(t_2);\cdots;\sigma(t_k)}_{\nu,\gamma}}\le
C_k \exp[-\frac{\max_{i,j}|t_i-t_j|}{\tau}]
\label{e:ssbound}
\end{equation}
for any $\gamma,\nu\in\Ss$ and any $t_1,t_2,\ldots,t_k$, provided that $k\ge2$.

We choose and fix $\taus$ such that $\taus\gg\tau$.
We assume that $T$ is large enough to satisfy $T\gg\taus$.

We shall evaluate \eqref{e:Skint} by decomposing the whole time interval to three regions $[0,\taus]$, $(\taus,T-\taus)$, and $[T-\taus,T]$ to see that 
the cumulant in question can be decomposed nicely as in \eqref{e:Skint4}.
Let us define the integral for the initial region
\begin{equation}
I^\mathrm{init}_{\nu}
=\int_0^{\taus} dt_1\int_0^{\taus}dt_2\cdots\int_0^{\taus}dt_k
\,\sbkt{\sigma(t_1);\sigma(t_2);\cdots;\sigma(t_k)}_{\nu,\gamma}
\label{e:Skint1}
\end{equation}
Although the definition contains $\gamma$, we see from the  the assumptions \eqref{e:ssbound} and $T\gg\taus\gg\tau$ that the $\gamma$ dependence can be neglected.
Likewise the integral for the final region
\begin{equation}
I^\mathrm{fin}_\gamma
=\int_{T-\taus}^T  dt_1\int_{T-\taus}^T dt_2\cdots\int_{T-\taus}^T dt_k
\,\sbkt{\sigma(t_1);\sigma(t_2);\cdots;\sigma(t_k)}_{\nu,\gamma}
\label{e:Skint2}
\end{equation}
should depend only on $\gamma$.
Finally the integral of the intermediate region is
\begin{equation}
I^\mathrm{int}=\int_{(t_1,\ldots,t_k)\in K}dt_1\cdots dt_k
\,\sbkt{\sigma(t_1);\sigma(t_2);\cdots;\sigma(t_k)}_{\nu,\gamma}
\label{e:Skint3}
\end{equation}
where $K\subset[0,T]^k$ is the region which remains to be integrated, i.e., the set in which $t_i\in(\taus,T-\taus)$ for at least one $i$ or $t_i\in[0,\taus]$, $t_j\in[T-\taus,T]$ for at least one pair $i$, $j$.
Again the assumptions \eqref{e:ssbound} and $\taus\gg\tau$ imply  that \eqref{e:Skint3} is independent of $\nu$, $\gamma$.
To see this, it suffices to note that the contribution to \eqref{e:Skint3} from $(t_1,\ldots,t_k)$ in which at least one of them satisfies $t_i\in[0,\taus/2]$ or $t_i\in[T-\taus/2,T]$ is negligible.
Of course these integrals sum up to be the desired cumulant as
\begin{equation}
\sbkt{\Theta^k}^\mathrm{c}_{\nu,\gamma}=I^\mathrm{init}_{\nu}+I^\mathrm{int}+I^\mathrm{fin}_{\gamma}.
\label{e:Skint4}
\end{equation}
When $T$ grows with $\taus$ fixed, the intermediate integral $I^\mathrm{int}$ grows linearly in $T$ while $I^\mathrm{init}_{\nu}$ and $I^\mathrm{fin}_{\gamma}$ remain unchanged.

By using the decomposition \eqref{e:Skint4}, the desired \eqref{e:SkSkSkSk} follows immediately.
The property of $\sbkt{(\Sex)^k}^\mathrm{c}_{\sss,\gamma}$ and 
$\sbkt{(\Sex)^k}^\mathrm{c}_{\gammas,\sss}$ stated below \eqref{e:finalfull} also follows.

\section{Steady state with particle flow}
\label{s:paricleflow}
In the present section, we discuss  representations for the stationary distribution of a  nonequilibrium steady state with a steady flow of particles.
Such nonequilibrium states can be modeled by a system attached to multiple particle baths (reservoirs) with different chemical potentials as in (4) of Fig.~\ref{f:123}.
As one may guess, the situation is very close to the problem of heat conduction, but one must consider the particle flow in addition to the energy flow.
Consequently the entropy production should be defined as in \eqref{e:THN}, which  takes into account the entropy production by transfer of particles as well as that by energy transfer.
With this modification, we get almost the same representations as in \eqref{e:mainN} and \eqref{e:mainN2}.

\subsection{States and Hamiltonians}
Let us refine our notation so that we can treat the situation where  particles move between different parts.

We again consider a classical system of many particles.
We assume for simplicity that all  particles are identical.
Note that particles are always distinguishable in classical mechanics.
As is always done in classical statistical mechanics, we introduce suitable combinatorial factors and relabeling so that to treat the particles as if they are indistinguishable.

Again the whole system consists of a ``system'' and $n$ particle baths (reservoirs).
Let $\calS^{(N)}$ be the phase space of the whole system with $N$ particles.
A state in $\calS^{(N)}$ is still written as $\Gamma=(\bsr_1,\ldots,\bsr_N;\bsp_1,\ldots,\bsp_N)$, but now the position $\bsr_l$ of the $l$-th particle may be in the system or in one of the $n$ reservoirs.
By $\Ns(\Gamma)$ and $N_i(\Gamma)$, we denote the numbers of the particles
 in the system and the $i$-th reservoir, respectively, in the state $\Gamma$.
One thus have $N=\Ns(\Gamma)+\sum_{i=1}^n N_i(\Gamma)$ if $\Gamma\in\calS^{(N)}$.
For a given $\Gamma$, we denote by $\Gs$ the state obtained
 by extracting the coordinates and momenta of those particles contained in the system.
 By $\Gamma_i$ we denote the state obtained by doing the same for the $i$-th reservoir.

As in \eqref{e:Hstat}, the Hamiltonian of the whole system is written as
\begin{equation}
\Hstat(\Gamma)=\Hs(\Gs)+\sum_{i=1}^n\{\Hb_i(\Gamma_i)+\Hc_i(\Gamma_i,\Gs)\},
\label{e:Hstat2}
\end{equation}
where $\Hs(\Gs)$, $\Hb_i(\Gamma_i)$, and $\Hc_i(\Gamma_i,\Gs)$ are the Hamiltonians for the system, the $i$-th reservoir, and the coupling between the $i$-th reservoir and the system, respectively.
We assume that all the Hamiltonians satisfy the time-reversal symmetry $H(\Gamma)=H(\Gamma^*)$.
We can again consider the standard choice as in \eqref{e:HsE}, \eqref{e:HbE}, and \eqref{e:HcE} (with suitable rearrangement of the labels of the  particles).

Again it is standard to consider the weak coupling limit, in which one neglects $\Hc_i(\Gamma_i,\Gs)$.
Let us remark that one may set exactly $\Hc_i(\Gamma_i,\Gs)=0$ and still have a meaningful model where the system and the reservoirs effectively interact with each other.
For this, we take a rather artificial model in which particles in the system do not interact with those in the reservoirs.
Particles can interact with each other in each reservoir or in the system.
Particles still move between the system and the reservoirs, thus generating (not necessarily weak) effective interactions between them.

We again write
\begin{equation}
H_i(\Gamma)=\Hb_i(\Gamma_i)+\Hc_i(\Gamma_i,\Gs).
\label{e:Hi2}
\end{equation}
By $\Te(\cdot)$, we denote the time evolution map from $t=0$ to $t=T$ determined by the pure Hamiltonian dynamics with the static Hamiltonian \eqref{e:Hstat2} .
If necessary one can further add a non-conservative force or an oscillating external force to the present setting as in the examples 2 and 3 of section~\ref{s:examples}.

\subsection{Initial state and the steady state}
We now define our initial state.
We assign the inverse temperature $\beta_i$ and the chemical potential $\mu_i$ to the $i$-th reservoir.

Let $\Ss^{(\Ns)}$ consist of  states $(\bsr_1^{(\mathrm{s})},\ldots,\bsr_{N_\mathrm{s}}^{(\mathrm{s})};\bsp_1^{(\mathrm{s})},\ldots,\bsp_{N_\mathrm{s}}^{(\mathrm{s})})$ where all $\bsr_j^{(\mathrm{s})}$ ($j=1,\ldots,\Ns$) are in the system.
In short,  $\Ss^{(\Ns)}$ is the state space of the system, where the labels of the particles happen to be $1,\ldots,\Ns$.

We take an arbitrary particle number $\Ns$ and a  state $\nu\in\Ss^{(\Ns)}$ of the system.
Then, as in \eqref{e:Pgamma}, we take the following initial distribution
in which the system is in the fixed state $\nu$ (after a possible relabeling) and the states of the reservoirs are distributed
 according to the grand canonical ensemble with the specified $\beta_i$ and $\mu_i$ as
\begin{equation}
P_{\nu}(\Gamma):=
\sum_\calL\delta(\Gs-\calL(\nu))\,\frac{1}{\Xi_{\nu}(\bsbeta,\bsmu)}\,
\exp\Bigl[-\sum_{i=1}^n\beta_i\,\{\Hb_i(\Gamma_i)+\Hc_i(\Gamma_i,\nu)-\mu_i\,N_i(\Gamma)\}\Bigr].
\label{e:Pgamma2}
\end{equation}
Note that we are considering an ensemble which contains various $N$.
We set $P_{\nu}(\Gamma)=0$ for $\Gamma\in\calS^{(N)}$ with $N<\Ns$.
Here $\calL$ is a map which changes the labels of the particles from $(1,2,\ldots,\Ns)$ to an arbitrary (ordered and non-overlapping) sequence chosen from $(1,2,\ldots,N)$.
In \eqref{e:Pgamma2}, $\calL$ is summed over $N!/(N-\Ns)!$ such relabelings.

The grand partition function $\Xi_\nu(\bsbeta,\bsmu)$ is determined by the normalization
\begin{equation}
\sum_{N=0}^\infty\int_{\Gamma\in\calS^{(N)}}\frac{\dG}{N!}\,P_{\nu}(\Gamma)=1.
\label{e:Xinorm}
\end{equation}
From \eqref{e:Pgamma2}, \eqref{e:Xinorm} and \eqref{e:Hi2}, we get\footnote{
What follows is a standard manipulation we always encounter when treating classical identical particles.
}
\begin{align}
\Xi_{\nu}(\bsbeta,\bsmu)&=
\sum_{N=\Ns}^\infty\sum_\calL\int_{\Gamma\in\calS^{(N)}}\frac{\dG}{N!}\,
\delta(\Gs-\calL(\nu))\,
\exp\Bigl[-\sum_{i=1}^n\beta_i\,\{H_i(\Gamma)-\mu_i\,N_i(\Gamma)\}\Bigr]
\notag
\intertext{Since the result of the integration does not depend on $\calL$, we have}
&=
\sum_{N=\Ns}^\infty\frac{N!}{(N-\Ns)!}\int_{\Gamma\in\calS^{(N)}}\frac{\dG}{N!}\,
\delta(\Gs-\nu)\,
\exp\Bigl[-\sum_{i=1}^n\beta_i\,\{H_i(\Gamma)-\mu_i\,N_i(\Gamma)\}\Bigr]
\ret
&=
\sum_{N'=0}^\infty\int\frac{d\Gamma'}{N'!}\,
\exp\Bigl[-\sum_{i=1}^n\beta_i\,\{\Hb_i(\Gamma'_i)+\Hc_i(\Gamma'_i,\nu)-\mu_i\,N_i(\Gamma')\}\Bigr],
\notag
\intertext{where $N'=N-\Ns$, and $\Gamma'$ is integrated over all the states $\Gamma'=(\bsr_{\Ns+1},\ldots,\bsr_N;\bsp_{\Ns+1},\ldots,\bsp_N)$ where all $\bsr_j$ are in the reservoirs.
We now decompose $N'$ as $N'=\sum_{i=1}^nN_i$ where $N_i$ is the number of particles in the $i$-th bath.
Then we further get
}
&=\sum_{N_1,\ldots,N_n=0}^\infty\frac{(\sum_iN_i)!}{\prod_iN_i!}
\int\frac{d\Gamma_1\cdots d\Gamma_n}{(\sum_iN_i)!}\exp\Bigl[-\sum_{i=1}^n\beta_i\,\{\Hb_i(\Gamma_i)+\Hc_i(\Gamma_i,\nu)-\mu_i\,N_i\}\Bigr],
\notag
\intertext{where (with a slight abuse of notation) $\Gamma_i$ is integrated over the states $(\bsr_{1},\ldots,\bsr_{N_i};\bsp_{1},\ldots,\bsp_{N_i})$ where all $\bsr_j$ are in the $i$-th reservoir.
The factor $(\sum_iN_i)!/\prod_iN_i!$ counts the ways of distributing the particles to the reservoirs.
This factorizes into a product of equilibrium grand partition functions (with the boundary condition $\nu$) as}
&=\prod_{i=1}^n\int\frac{d\Gamma_i}{N_i!}
\exp[-\beta_i\,\{\Hb_i(\Gamma_i)+\Hc_i(\Gamma_i,\nu)-\mu_i\,N_i\}].
\label{e:Xinu}
\end{align}

We again define the corresponding grand partition function without coupling terms as
\begin{equation}
\tilde{\Xi}(\bsbeta,\bsmu):=\prod_{i=1}^n\int\frac{d\Gamma_i}{N_i!}
\exp[-\beta_i\,\{\Hb_i(\Gamma_i)-\mu_i\,N_i\}],
\label{e:tXinu}
\end{equation}
and the ``free energy'' of the coupling as
\begin{equation}
\phic({\nu}):=-\log\frac{\Xi_{\nu}(\bsbeta,\bsmu)}{\tilde{\Xi}(\bsbeta,\bsmu)}.
\label{e:Xgf}
\end{equation}
As before, $\phic({\nu})$ is an equilibrium quantity which vanishes in the weak coupling limit.

We shall fix the final condition for the system by taking an arbitrary $\Ns'$ and the corresponding state $\gamma\in\Ss^{(\Ns')}$ of the system.
We define
\begin{equation}
\rho_\nu(\gamma):=
\sum_{N=0}^\infty\int_{\Gamma\in\calS^{(N)}}\frac{\dG}{N!}\,P_{\nu}(\Gamma)\,
\sum_{\calL'}\delta[\ofs{\Te(\Gamma)}-\calL'(\gamma)],
\label{e:rng2}
\end{equation}
where $\calL'$ is summed over all possible relabelings as in \eqref{e:Pgamma2}.
Note that $\rho_\nu(\gamma)$ is normalized as
\begin{equation}
\sum_{\Ms=0}^\infty\int_{\gamma\in\Ss^{(\Ms)}}\frac{d\gamma}{\Ms!}\,\rho_\nu(\gamma)
=1
\label{e:rngnorm}
\end{equation}
for any $\nu$.
To see this we observe that 
\begin{equation}
\int_{\gamma\in\Ss^{(\Ms)}}d\gamma\,\sum_{\calL'}\delta[\Gs-\calL'(\gamma)]
=
\begin{cases}
\Ms!&\text{if $\Gs$ contains $\Ms$ particles}\\
0&\text{otherwise}
\end{cases}
\label{e:rngnorm0}
\end{equation}
for any $\Gamma$, and recall the normalization \eqref{e:Xinorm}.

For an arbitrary function $f(\Gamma)$ (where $\Gamma$ is an element of $\calS^{(N)}$ with variable $N$), we define the average
\begin{equation}
\sbkt{f}_{\nu,\gamma}=
\frac{1}{\rho_\nu(\gamma)}
\sum_{N=0}^\infty\int_{\Gamma\in\calS^{(N)}}\frac{\dG}{N!}\,f(\Gamma)\,P_{\nu}(\Gamma)\,
\sum_{\calL'}\delta[\ofs{\Te(\Gamma)}-\calL'(\gamma)],
\label{e:fng2}
\end{equation}
in which the initial and the final states of the system are fixed.

Again we assume that the reservoirs are so large and $T$ is so large that we can set
\begin{equation}
\rho_\nu(\gamma)=\rhoss(\gamma)
\label{e:rssrg2}
\end{equation}
for any $\nu$, where $\rhoss(\gamma)$ is the  stationary distribution for the unique  nonequilibrium steady state.
Partially conditioned averages $\sbkt{f}_{\nu,\sss}$ and $\sbkt{f}_{\sss,\gamma}$ are defined as \eqref{e:fgs} and \eqref{e:fsg}, respectively.

\subsection{Entropy production and the representation of $\rhoss(\gamma)$}
We now define the entropy production in the trajectory $\Gamma\rightsquigarrow\Te(\Gamma)$ as
\begin{equation}
\Theta(\Gamma):=\sum_{i=1}^n\left\{\,
 \beta_i\,[H_i(\Te(\Gamma))-H_i(\Gamma)]
-\beta_i\mu_i\,[N_i(\Te(\Gamma))-N_i(\Gamma)]
\right\},
\label{e:THN}
\end{equation}
where $H_i(\Gamma)$ is the Hamiltonian for the $i$-th reservoir defined in \eqref{e:Hi2}.
This is the total entropy production which takes into account the transfer of particles as well as that of energy.

For an arbitrary function $f(\Gamma)$, we can show the identity
\begin{equation}
\Xi_{\nu}\,\rho_{\nu}(\gamma)\,\sbkt{f\,e^{-\Theta/2}}_{\nu,\gamma}
=
\Xi_{\gammas}\,\rho_{\gammas}(\nus)\,\sbkt{f^\dagger\,e^{-\Theta/2}}_{\gammas,\nus},
\label{e:Xrf}
\end{equation}
for any $\Ns$, $\Ms$ and any $\nu\in\Ss^{(\Ns)}$, $\gamma\in\Ss^{(\Ms)}$.
This identity precisely corresponds to \eqref{e:ZrfZrf}.
To derive \eqref{e:Xrf}, we first note that the definitions \eqref{e:Pgamma2} and \eqref{e:fng2} imply
\begin{align}
\Xi_{\nu}\,\rho_{\nu}(\gamma)\,\sbkt{f}_{\nu,\gamma}&=
\sum_{N=0}^\infty\int_{\Gamma\in\calS^{(N)}}\frac{\dG}{N!}\,f(\Gamma)
\,\exp\bigl[-\sum_{i=1}^n\beta_i\{H_i(\Gamma)-\mu_i\,N_i(\Gamma)\}\bigr]\times
\ret
&\hspace{2cm}\times\sum_{\calL,\calL'}\delta(\Gs-\calL(\nu))\,\delta[\ofs{\Te(\Gamma)}-\calL'(\gamma)],
\label{e:Xrf00}
\end{align}
which corresponds to \eqref{e:Zrf00}.
Then  \eqref{e:Xrf} can be derived in essentially the same way as  \eqref{e:ZrfZrf}.


By starting from \eqref{e:Xrf} and repeating the derivation in sections~\ref{s:repder} and \ref{s:OE}, we get the desired representation
\begin{equation}
\pss(\gamma)=\varphi^{(0)}+\phic({\gamma})+
\frac{1}{2}\{\sbkt{\Sex}_{\gammas,\sss}-\sbkt{\Sex}_{\sss,\gamma}\}
+O(\epsilon^3),
\label{e:mainN}
\end{equation}
where we wrote $\pss(\gamma):=-\log\rhoss(\gamma)$.
Here $\varphi^{(0)}$ is a constant independent of $\gamma$, and $\phic({\gamma})$ is the free energy for coupling defined in \eqref{e:Xgf}.
We recall that there are sensible models in which $\phic({\gamma})$ are vanishing.
The excess entropy production is of course defined as $\Theta_\mathrm{ex}(\Gamma)=\Theta(\Gamma)-\barsigma\,T$, where $\barsigma$ is the entropy production rate.
See section~\ref{s:EP}.

The representation \eqref{e:mainN} may be most useful in the situation where all the reservoirs have the same inverse temperature.
Let us set $\beta_i=\beta$ for $i=1,\ldots,n$, and note that the energy conservation implies
\begin{equation}
\sum_{i=1}^n\beta\,\{H_i(\Te(\Gamma))-H_i(\Gamma)\}=\beta\,\{\Hs(\Gs)-\Hs((\Te(\Gamma))_\mathrm{s})\}.
\label{e:ECN}
\end{equation}
Then the entropy production \eqref{e:THN} is written as
\begin{equation}
\Theta(\Gamma)=\beta\,\{\Hs(\Gs)-\Hs((\Te(\Gamma))_\mathrm{s})\}+\Psi(\Gamma),
\label{e:THHP}
\end{equation}
where
\begin{equation}
\Psi(\Gamma)=-\beta\sum_{i=1}^n\mu_i\,\{N_i(\Te(\Gamma))-N_i(\Gamma)\}
\label{e:Psi}
\end{equation}
is the entropy production due to the transfer of particles.
Substituting \eqref{e:THHP} into the representation \eqref{e:mainN}, we get 
\begin{equation}
\pss(\gamma)=\varphi^{(1)}+\phic(\gamma)+\beta\,\Hs(\gamma)+
\frac{1}{2}\{\sbkt{\Psi_\mathrm{ex}}_{\gammas,\sss}-\sbkt{\Psi_\mathrm{ex}}_{\sss,\gamma}\}
+O(\epsilon^3).
\label{e:mainN2}
\end{equation}
The constant is given by $\varphi^{(1)}=\varphi^{(0)}-\beta\,\bar{\Hs}$ where $\bar{\Hs}=\sum_{\Ns=0}^\infty\int_{\gamma\in\Ss^{(\Ns)}}(d\gamma/\Ns!)\,\rhoss(\gamma)\,\Hs(\gamma)$ is the expectation value of the energy in the steady state.
The excess quantity is again defined as $\Psi_\mathrm{ex}(\Gamma)=\Psi(\Gamma)-\barsigma\,T$.

It is an easy exercise to rewrite, as we have done in section~\ref{s:canonical}, the representations \eqref{e:mainN} and \eqref{e:mainN2} in the forms where the contributions from the equilibrium grand canonical distribution becomes manifest.


\begin{thebibliography}{10}

\bibitem{ST}
S. Sasa and H. Tasaki,
J. Stat. Phys. {\bf 125}, 125 (2006).



\bibitem{KuboTodaHashitsume85}
R. Kubo, M. Toda, and N. Hashitsume,
{\em Statistical Physics II}\/ (Springer, 1985).

\bibitem{McLennan90}
J. A. Mclennan,
{\em Introduction to nonequilibrium statistical mechanics}
(Prentice Hall, 1990). 

\bibitem{Zubarev74}
D. N. Zubarev, 
{\em Nonequilibrium statistical thermodynamics}
(Consultants Bureau, 1974). 

\bibitem{KawasakiGunton73}
K. Kawasaki and J. D. Gunton, 
Phys. Rev. A {\bf 8}, 2048--2064 (1973). 


\bibitem{ECM}
D. J. Evans,  E. G. D. Cohen,  and G. P. Morris,
{Phys. Rev. Lett.} 
\textbf{71},  2401 (1993).

\bibitem{GC}
G. Gallavotti and E. G. D. Cohen,
{ Phys. Rev. Lett.}
\textbf{74},   2694 (1995).

\bibitem{Kurchan}
J. Kurchan,  
{J. Phys. A: Math. Gen.}
\textbf{31},   3719 (1998).

\bibitem{Maes}
C. Maes, 
{J. Stat. Phys.}  
\textbf{95},  367 (1999).

\bibitem{Crooks} 
{G. E. Crooks},
{Phys. Rev. E}
\textbf{61},
{2361} {(2000)}.

\bibitem{J}
C. Jarzynski,
J. Stat. Phys. {\bf 98}, 77 (2000).

\bibitem{MN}
C. Maes and K. Neto\v{c}n\'{y}, J. Stat. Phys. {\bf 110}, 269
(2003).


\bibitem{KNPRL}
T. S. Komatsu and N. Nakagawa,
Phys. Rev. Lett. {\bf 100}, 030601 (2008) archived as 0708.3158.

\bibitem{KNSTshort}
T. S. Komatsu, N. Nakagawa, S. Sasa, and H. Tasaki,
preprint archived as 0711.0246; to appear in Phys. Rev. Lett.

\bibitem{OP}
{Y. Oono and M. Paniconi},
{Prog. Theor. Phys. Suppl.}
\textbf{130}, 29 (1998).


\bibitem{HS}
{T. Hatano and S. Sasa},
{Phys. Rev. Lett.}
\textbf{86}, 3463 
{(2001)}.




\end{thebibliography}
\end{document}